%% file: main.tex
\documentclass[10pt,journal,compsoc]{IEEEtran}
\usepackage[T1]{fontenc}
\usepackage{booktabs} 
\usepackage{tabularx}
\usepackage{graphicx}
\usepackage{url}
\usepackage{mdwlist}
\usepackage{subcaption}
\usepackage{hyperref}
\hypersetup{colorlinks=false,linkcolor=blue,urlcolor=blue,citecolor=red}
\usepackage{amsfonts}
\usepackage[labelfont=bf,textfont=bf]{caption}
\usepackage{multirow}
\usepackage{array}
\usepackage[linesnumbered,ruled]{algorithm2e}
\SetKwComment{Comment}{$\triangleright$\ }{}
\usepackage{xcolor}
\usepackage{amsmath}
\usepackage{amssymb}
\usepackage[flushleft]{threeparttable}
\usepackage{enumitem}
\usepackage{balance}

\newcommand{\ie}{\emph{i.e., }}
\newcommand{\eg}{\emph{e.g., }}
\newcommand{\etal}{\emph{et al.}}
\newcommand{\st}{\emph{s.t. }}

\newcommand{\wrt}{\emph{w.r.t. }}
\newcommand{\aka}{\emph{a.k.a. }}

\ifCLASSOPTIONcompsoc
  \usepackage[nocompress]{cite}
\else
  \usepackage{cite}
\fi
\ifCLASSINFOpdf
\else
\fi
\begin{document}

\title{Adversarial Training Towards Robust \\Multimedia Recommender System}

\author{Jinhui~Tang,~\IEEEmembership{Senior~Member,~IEEE,~}%
        Xiaoyu~Du, %
        Xiangnan~He, %
        Fajie~Yuan, %
        Qi~Tian,~\IEEEmembership{Fellow,~IEEE,~}%
	and~Tat-Seng~Chua
\IEEEcompsocitemizethanks{\IEEEcompsocthanksitem Jinhui Tang is with the Nanjing University of Science and Technology, Nanjing, Jiangsu, China, 210094.\protect\\
E-mail: jinhuitang@njust.edu.cn
\IEEEcompsocthanksitem Xiaoyu Du is with the University of Electronic Science and Technology of China, Chengdu, Sichuan, China, 610054.\protect\\
E-mail: duxy.me@gmail.com
\IEEEcompsocthanksitem Xiangnan He is with the University of Science and Technology of China, Hefei, Anhui, China, 230031. \protect\\
E-mail: xiangnanhe@gmail.com
\IEEEcompsocthanksitem Fajie Yuan is with the Platform and Content Group (PCG) of Tencent Shenzhen, Guangdong, China, 518057.\protect\\
E-mail: fajieyuan@tencent.com
\IEEEcompsocthanksitem Qi Tian is with the Huawei Noah's Ark Laboratory, Huawei, China.\protect\\
E-mail: tian.qi1@huawei.com
\IEEEcompsocthanksitem Tat-Seng Chua is with the National University of Singapore, Singapore, 117417.\protect\\
E-mail: dcscts@nus.edu.sg
}
\thanks{Xiangnan He is the corresponding author. Manuscript received Aug 1, 2018.}
}

\markboth{IEEE Transactions on Knowledge and Data Engineering}%
{Tang \MakeLowercase{\textit{et al.}}: Adversarial Training Towards Robust Multimedia Recommender System}

\IEEEtitleabstractindextext{%
\begin{abstract}
With the prevalence of multimedia content on the Web, developing recommender solutions that can effectively leverage the rich signal in multimedia data is in urgent need. Owing to the success of deep neural networks in representation learning, recent advance on multimedia recommendation has largely focused on exploring deep learning methods to improve the recommendation accuracy. To date, however, there has been little effort to investigate the robustness of multimedia representation and its impact on the performance of multimedia recommendation. 
	
	In this paper, we shed light on the robustness of multimedia recommender system. Using the state-of-the-art recommendation framework and deep image features, we demonstrate that the overall system is not robust, such that a small (but purposeful) perturbation on the input image will severely decrease the recommendation accuracy. This implies the possible weakness of multimedia recommender system in predicting user preference, and more importantly, the potential of improvement by enhancing its robustness. 
To this end, we propose a novel solution named \textit{Adversarial Multimedia Recommendation} (AMR), which can lead to a more robust multimedia recommender model by using adversarial learning. The idea is to train the model to defend an adversary, which adds perturbations to the target image with the purpose of decreasing the model's accuracy. We conduct experiments on two representative multimedia recommendation tasks, namely, image recommendation and visually-aware product recommendation. Extensive results verify the positive effect of adversarial learning and demonstrate the effectiveness of our AMR method. Source codes are available in \url{https://github.com/duxy-me/AMR}. 
\end{abstract}

}

\maketitle

\IEEEdisplaynontitleabstractindextext

%
\IEEEpeerreviewmaketitle


\input{1_introduction}
\input{2_preliminary}
\input{3_method}
\input{5_experiment}
\input{4_related}
\input{6_conclusion}



\ifCLASSOPTIONcaptionsoff
  \newpage
\fi



%
\bibliographystyle{IEEEtran}
%

\bibliography{reference}

%

\begin{IEEEbiography}[{\includegraphics[width=1in,height=1.25in,clip,keepaspectratio]{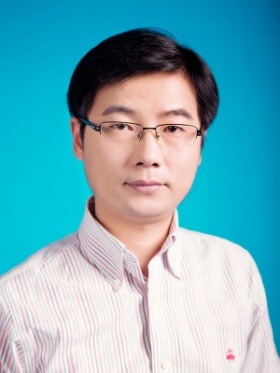}}]{Jinhui Tang} (M'08-SM'14) received the B.Eng. and Ph.D. degrees from the University of Science and Technology of China, Hefei, China, in 2003 and 2008, respectively. He is currently a Professor with the School of Computer Science and Engineering, Nanjing University of Science and Technology, Nanjing, China. From 2008 to 2010, he was a Research Fellow with the School of Computing, National University of Singapore, Singapore. He has authored over 150 papers in top-tier journals and conferences. His current research interests include multimedia analysis and search, computer vision, and machine learning. Dr. Tang was a recipient of the best paper awards in ACM MM 2007, PCM 2011, and ICIMCS 2011, the Best Paper Runner-up in ACM MM 2015, and the best student paper awards in MMM 2016 and ICIMCS 2017. He has served as an Associate Editor for the IEEE TKDE, IEEE TNNLS, and IEEE TCSVT.
\end{IEEEbiography}

\begin{IEEEbiography}[{\includegraphics[width=1in,height=1.25in,clip,keepaspectratio]{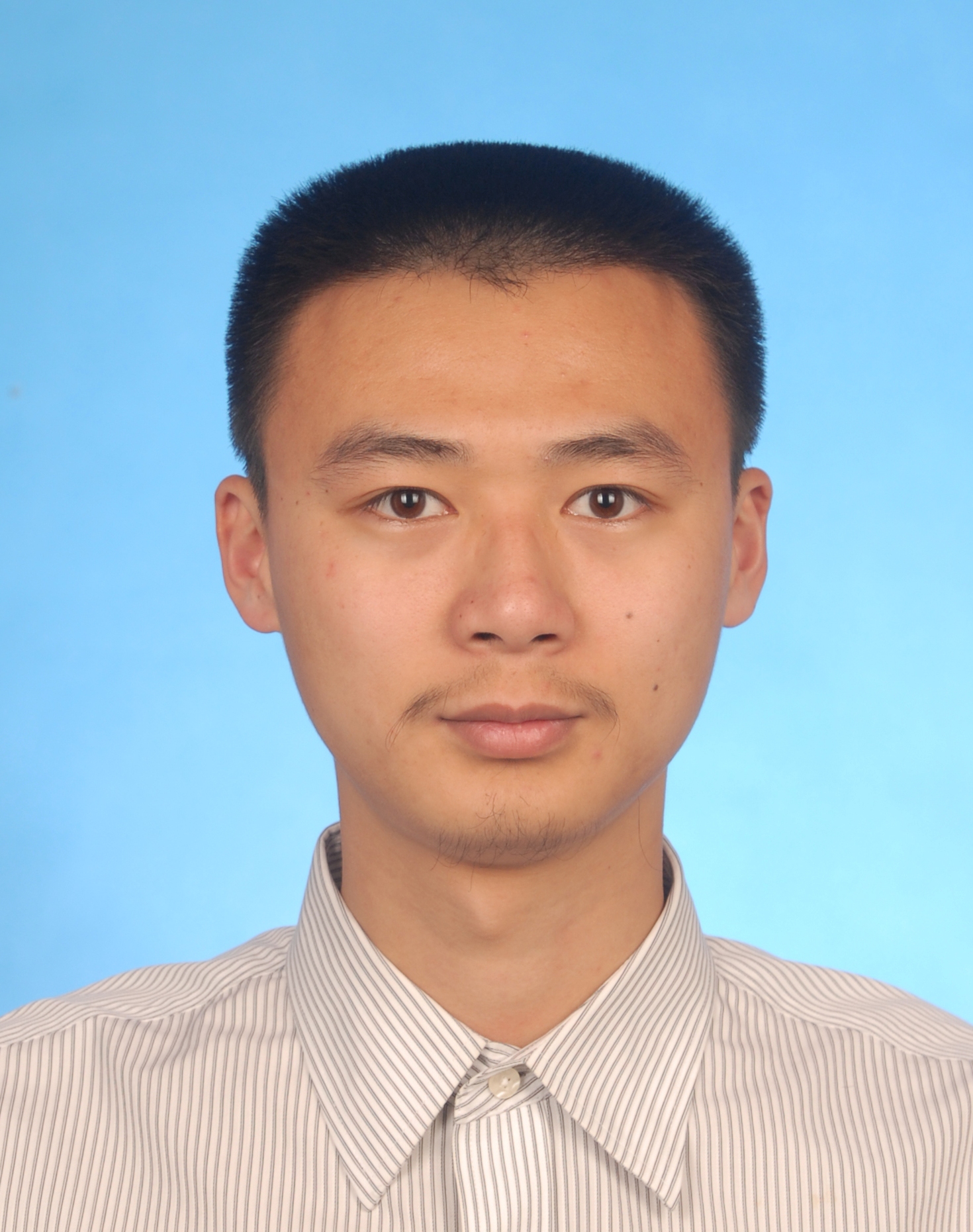}}]{Xiaoyu Du}
is currently a lecturer in the School of Software Engineering of Chengdu University of Information Technology, Chengdu, 
a visiting scholar in the NeXT++ of National University of Singapore,
and a Ph.D. candidate of University of Electronic Science and Technology of China, Chengdu. 
He received his M.E. degree in computer software and theory in 2011 and B.S. degree in computer science and technology in 2008, both from Beijing Normal University, Beijing. 
His research interests include information retrieval, computer vision, and machine learning.
\end{IEEEbiography}

\begin{IEEEbiography}[{\includegraphics[width=1in,height=1.25in,clip,keepaspectratio]{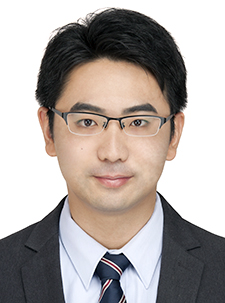}}]{Xiangnan He} is currently a professor with the  University of Science and Technology of China (USTC). He received his Ph.D. in Computer Science from National University of Singapore (NUS) in 2016, and did postdoctoral research in NUS until 2018. His research interests span information retrieval, data mining, and multi-media analytics. He has over 50 publications appeared in several top conferences such as SIGIR, WWW, and MM, and journals including TKDE, TOIS, and TMM. His work on recommender systems has received the Best Paper Award Honourable Mention in WWW 2018 and ACM SIGIR 2016. Moreover, he has served as the PC member for several top conferences including SIGIR, WWW, MM, KDD etc., and the regular reviewer for journals including TKDE, TOIS, TMM, TNNLS etc.
\end{IEEEbiography}

\begin{IEEEbiography}[{\includegraphics[width=1in,height=1.25in,clip,keepaspectratio]{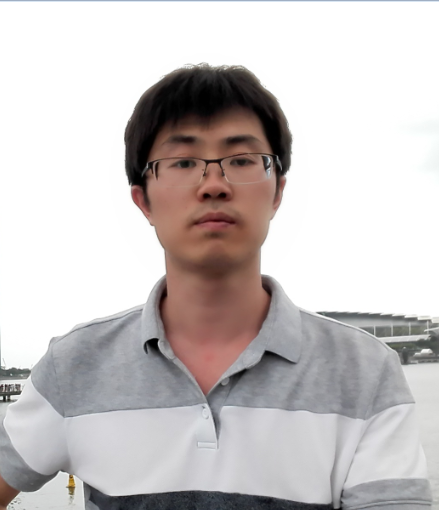}}]{Fajie Yuan}
is now a senior researcher in the PCG (Platform \& Content) group of Tencent Inc., China. He received his Ph.D. in the School of Computing Science at the University of Glasgow in 2018. His research interests span recommender system (RS), natural language processing (NLP) and machine learning (ML). His work have appeared in several top-tier conferences, including UAI, WSDM, IJCAI, and ACL, and he has won the best student paper award in ICTAI 2016.

\end{IEEEbiography}
\begin{IEEEbiography}[{\includegraphics[width=1in,height=1.25in,clip,keepaspectratio]{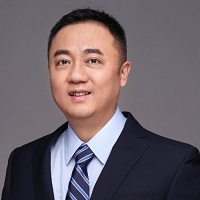}}]{Qi Tian} is a Chief Scientist of Computer Vision in Huawei Noah's Ark Laboratory, a Full Professor in the Department of Computer Science, The University of Texas at San Antonio (UTSA), a Changjiang Chaired Professor of the Ministry of Education, an Overseas Outstanding Youth, an Overseas Expert by the Chinese Academy of Sciences, and an IEEE Fellow. He was also a Visiting Chaired Professor in the Center for Neural and Cognitive Computation, Tsinghua University, a Lead Researcher in the Media Computing Group at Microsoft Research Asia (MSRA). He received his B.E. in Electronic Engineering from Tsinghua University and then he went to the United States and received his Ph.D. in ECE from the University of Illinois at Urbana-Champaign (UIUC).
\end{IEEEbiography}

\begin{IEEEbiography}[{\includegraphics[width=1in,height=1.25in,clip,keepaspectratio]{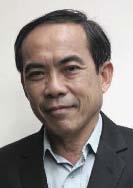}}]{Tat-Seng Chua}
is the KITHCT Chair Professor at the School of Computing, National University of Singapore. He was the Acting and Founding Dean of the School during 1998-2000. Dr Chuas main research interest is in multimedia information retrieval and social media analytics. In particular, his research focuses on the extraction, retrieval and question-answering (QA) of text and rich media arising from the Web and multiple social networks. He is the co-Director of NExT, a joint Center between NUS and Tsinghua University to develop technologies for live social media search. Dr Chua is the 2015 winner of the prestigious ACM SIGMM award for Outstanding Technical Contributions to Multimedia Computing, Communications and Applications. He is the Chair of steering committee of ACM International Conference on Multimedia Retrieval (ICMR) and Multimedia Modeling (MMM) conference series. Dr Chua is also the General Co-Chair of ACM Multimedia 2005, ACM CIVR (now ACM ICMR) 2005, ACM SIGIR 2008, and ACM Web Science 2015. He serves in the editorial boards of four international journals. Dr. Chua is the co-Founder of two technology startup companies in Singapore. He holds a PhD from the University of Leeds, UK.
\end{IEEEbiography}




\end{document}

%% file: 1_introduction.tex
\IEEEraisesectionheading{\section{Introduction}\label{sec:introduction}}
\IEEEPARstart{R}{ecommender} system plays a central role in user-centric online services, such as E-commerce, media-sharing, and social networking sites. By providing personalized content suggestions to users, recommender system not only can alleviate the information overload issue and improve user experience, but also can increase the profit for content providers through increasing the traffic. Thus many research efforts have been devoted to advance recommendation technologies, which have become an attractive topic of research in both academia and industry in the recent decade~\cite{he2018nais,yuan2016lambdafm,wang2018exploring,lian2018scalable,wang2017learning}. On the other hand, multimedia data becomes prevalent on the current Web. For example, products are usually associated with images to attract customers in E-commerce sites~\cite{yu2018aesthetic}, and users usually post images or micro-videos to interact with their friends in social media sites~\cite{Chen:2016,Zhang:2016:microvideo}. Such multimedia content contains rich visually-relevant signal that can reveal user preference~\cite{wang2015visual}, providing opportunities to improve recommender systems that are typically based on collaborative filtering on user behavior data only~\cite{rendle2009bpr,NCF}.

Early multimedia recommendation works have largely employed annotated tags~\cite{JustClick,chen2012personalized} or low-level representations~\cite{su2011efficient} such as color-based features and texture features like SFIT, to capture the semantics of multimedia content. Owing to the success of deep neural networks (DNNs) in learning representations~\cite{he2016deep}, recent advance on multimedia recommendation has shifted to integrating deep multimedia features into recommender model~\cite{ACF,Chen:2017:PKF,ICCV15,Oord:2013:DCM,cheng2018mmalfm}. For example, in image-based recommendation, a typical paradigm is to project the CNN features of image into the same latent space as that of users~\cite{VBPR,ICCV15}, or simultaneously learn image representation and recommender model~\cite{lei2016comparative}. 

Although the use of DNNs to learn multimedia representation leads to better recommendation performance than manually crafted features, we argue that a possible downside is that the overall system becomes \textbf{less robust}. As have shown in several previous works~\cite{AML_2014,AML_2015,moosavi2016universal}, many state-of-the-art DNNs are vulnerable to adversarial attacks. Taking the image classification task as an example, by applying small but intentionally perturbations to well-trained images from the dataset, these DNN models output wrong labels for the images with high confidence. This implies that the image representations learned by DNNs are not robust, which further, may negatively affect downstream applications based on the learned representations. 

\begin{figure}
	\includegraphics[width=\linewidth]{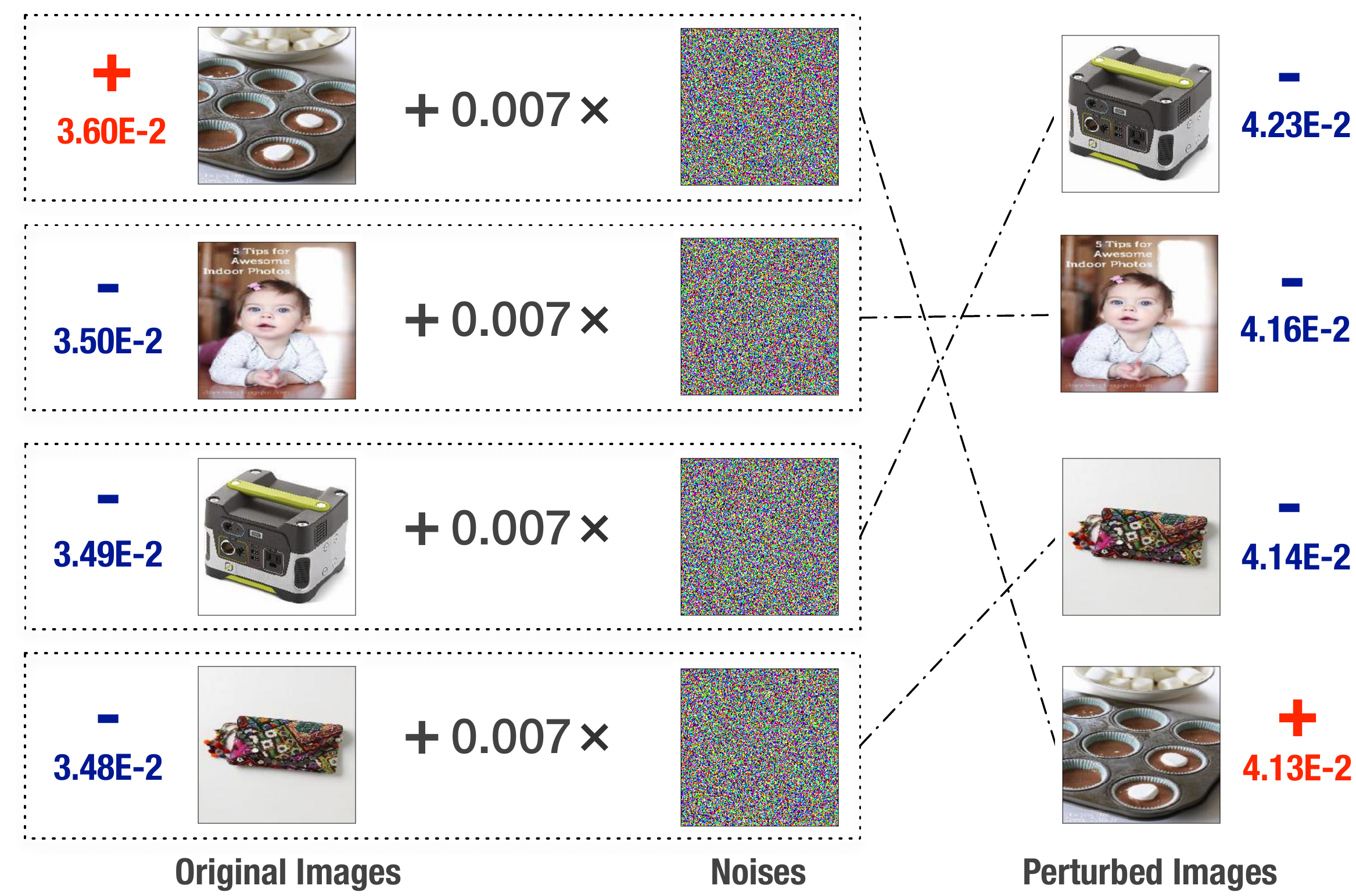}
	\caption{An example on how small perturbations on images would have a profound impact on the recommendation results. We sampled a user, one interacted image (in red ``+'') and three non-interacted images (in blue ``-'') by the user. The number besides each image is the ranking score generated by VBPR~\cite{VBPR} before (left) and after (right) perturbations. By adding small perturbations with the scale of $\epsilon=0.007$, the positive image is ranked much lower than before, even though the difference of perturbed images can be hardly perceived by human. Here the perturbations are generated by the fast gradient sign method~\cite{AML_2015}.}\label{fig:examples}\vspace{-10pt}
\end{figure}

Figure \ref{fig:examples} shows an illustrative example on how the lack of robustness affects the recommendation results. We first trained the Visual Bayesian Personalized Ranking (VBPR) method~\cite{VBPR} on a Pinterest dataset; VBPR is a state-of-the-art visually-aware recommendation method, and we used the ResNet-50~\cite{he2016deep} to extract image features for it. We then sampled a user $u$, showing her interacted image in the testing set (\ie the top-left image with sign ``+'') and three non-interacted images (\ie the bottom-left three images with sign ``-''). From the prediction scores of VBPR (\ie the numbers beside images), we can see that, originally, VBPR successfully ranks the positive image higher than other negative images for the user. However, after applying adversarial perturbations to these images, even though the perturbation scale is very small ($\epsilon=0.007$) \st human can hardly perceive the change on the perturbed images, VBPR outputs very different prediction scores and fails to rank the positive image higher than other negative images. This example demonstrates that adversarial perturbations for DNNs would have a profound impact on the downstream recommender model, making the model less robust and weak in generalizing to unseen predictions. 

In this paper, we enhance the robustness of multimedia recommender system and thus its generalization performance by performing adversarial learning~\cite{AML_2017}. With VBPR as the main recommender model, we introduce an adversary which adds perturbations on multimedia content with the aim of maximizing the VBPR loss function. We term our method as \textit{Adversarial Multimedia Recommendation} (AMR), which can be interpreted as playing a minimax game --- the perturbations are learned towards maximizing the VBPR loss, whereas the model parameters are learned towards minimizing both the VPBR loss and the adversary's loss. Through this way, we can enhance model robustness to adversarial perturbations on the multimedia content, such that the perturbations have a smaller impact on the model's prediction. 
To verify our proposal, we conduct experiments on two public datasets, namely, the Pinterest image data~\cite{ICCV15} and Amazon product data~\cite{VBPR}. Empirical results demonstrate the positive effect of adversarial learning and the effectiveness of our AMR method for multimedia recommendation.  

We summarize the main contributions of this work as follows.
\begin{enumerate}
  \item This is the first work to emphasize the vulnerability issue of state-of-the-art multimedia recommender systems due to the use of DNNs for feature learning. 
  \item A novel method is proposed to train a more robust and effective recommender model by using the recent developments on adversarial learning. 
  \item Extensive experiments are conducted on two representative multimedia recommendation tasks of personalized image recommendation and visually-aware product recommendation to verify our method. 
\end{enumerate}

The remainder of the paper is organized as follows. We first provide some preliminaries in Section~\ref{sec:preliminary}, and then elaborate our proposed method in Section~\ref{sec:method}. We present experimental results in Section~\ref{sec:experiment} and review related literature in Section~\ref{sec:related}. Finally, we conclude this paper and discuss future directions in Section~\ref{sec:conclusion}. 

%% file: 2_preliminary.tex
\section{Preliminaries} \label{sec:preliminary}
This section provides some technical background to multimedia recommendation. We first recapitulate the Latent Factor Model (LFM), which is the most widely used recommender model in literature~\cite{fastMF}. We then introduce the Visual Bayesian Personalized Ranking (VBPR)~\cite{VBPR}, which is a state-of-the-art method for multimedia recommendation, and we use it as AMR's building block. 

\subsection{Latent Factor Model}
The key of recommendation is to estimate the preference of a user on an item. The paradigm of LFM is to describe a user (and an item) as a vector of latent factors, \aka \textit{latent vector}; then the preference score is estimated as the inner product of the user latent vector and item latent vector. Formally, let $u$ denote a user, $i$ denote an item, and $\hat{y}_{ui}$ denote the estimated preference score of $u$ on $i$. Then the predictive model of LFM can be abstracted as:
\begin{equation}
\hat{y}_{ui} = <f_U(u), f_I(i)>, 
\end{equation}
where $f_U$ denotes the function that projects a user to the latent space, \ie $f_U(u)$ denotes the latent vector for user $u$; similar semantics apply to $f_I$, the notation of item side. 

For a LFM, the design of function $f_U$ and $f_I$ plays a crucial role on its performance, whereas the design is also subjected to the availability of the features to describe a user and an item. In the simplest case, when only the ID information is available, a common choice is to directly associate a user (and an item) with a vector, \ie $f_U(u)=\textbf{p}_u$ and $f_I(i)=\textbf{q}_i$, where $\textbf{p}_u\in \mathbb{R}^K$ and $\textbf{q}_i\in \mathbb{R}^K$ are also called as the \textit{embedding vector} for user $u$ and item $i$, respectively, and $K$ denotes the embedding size. This instantiation is known as the matrix factorization (MF) model~\cite{fastMF}, a simple yet effective model for the collaborative filtering task. 

Targeting at multimedia recommendation, $f_I$ is typically designed to incorporate content-based features, so as to leverage the visual signal of multimedia item. For example, Geng \etal~\cite{ICCV15} defines it as $f_I(i) = \textbf{E} \textbf{c}_i$, where $ \textbf{c}_i\in \mathbb{R}^{4096}$ denotes the deep image features extracted by AlexNet~\cite{krizhevsky2012imagenet}, and $\textbf{E}\in \mathbb{R}^{K\times 4096}$ transforms the image features to the latent space of LFM. A side benefit of such content-based modeling is that the item cold-start issue can be alleviated, since for out-of-sample items, we can still obtain a rather reliable latent vector from its content features. 
Besides this straightforward way to incorporate multimedia content, other more complicated operations have also been developed. For example, the Attentive Collaborative Filtering (ACF) model~\cite{ACF} uses an attention network to discriminate the importance of different components of a multimedia item, such as the regions in an image and frames of a video. 

Owing to the strong generalization ability of LFM in predicting unseen user-item interactions, LFM is recognized as the most effective model for personalized recommendation~\cite{ACF}. As such, we build our adversarial recommendation method upon LFM, more specifically VBPR --- an instantiation of LFM for multimedia recommendation. Next, we describe the VBPR method.  

\subsection{Visual Bayesian Personalized Ranking}
It is arguable that a user would not buy a new clothing product from Amazon without seeing it in person, so the visual appearance of items plays an important role in user preference prediction. 
VBPR is designed to incorporate such visual signal into the learning of user preference from implicit feedback~\cite{VBPR}. To be specific, its predictive model is formulated as:
\begin{equation}\label{eq:vbpr}
\hat{y}_{ui} = \textbf{p}_u^T \textbf{q}_i + \textbf{h}_u^T (\textbf{E}\textbf{c}_i),
\end{equation}
where the first term $\textbf{p}_u^T \textbf{q}_i$ is same as MF to model the collaborative filtering effect, and the second term $\textbf{h}_u^T (\textbf{E}\textbf{c}_i)$ models user preference based on the item's image. Specifically, $\textbf{p}_u\in \mathbb{R}^K$ ($\textbf{q}_i\in \mathbb{R}^K$)  denotes the ID embedding for user $u$ (item $i$), $\textbf{h}_u\in \mathbb{R}^K$ is $u$'s embedding in the image latent space, $\textbf{c}_i\in \mathbb{R}^{D}$ denotes the visual feature vector for item $i$ (which is extracted by AlexNet), and $\textbf{E}\in \mathbb{R}^{K\times D}$ converts the visual feature vector to latent space. The $K$ is a hyper-parameter and the $D$ is 4096 if using AlexNet. We can interpret this model as a LFM by defining $f_{U}(u)=[\textbf{p}_u, \textbf{h}_u]$ and $f_{I}(i)=[\textbf{q}_i, \textbf{E}\textbf{c}_i]$, where $[\cdot,\cdot]$ denotes vector concatenation. 
Note that in Equation (\ref{eq:vbpr}), we have only included the key terms on the interaction prediction in VBPR and omitted other bias terms for clarity.

To estimate model parameters, VBPR optimizes the BPR pairwise ranking loss~\cite{rendle2009bpr} to tailor the model for implicit interaction data such as purchases and clicks. The assumption is that interacted user-item pairs should be scored higher than the non-interacted pairs by the model. To implement this assumption, for each observed interaction $(u,i)$, BPR maximizes the margin between it and its unobserved counterparts. The objective function to minimize is:
\begin{equation}\label{eq:bpr}
L_{BPR} = \sum_{(u,i,j)\in \mathcal{D}} -\ln \sigma(\hat{y}_{ui} - \hat{y}_{uj}) + \beta ||\Theta||^2,
\end{equation}
where $\sigma(\cdot)$ is the sigmoid function, $\beta$ controls the strength of $L_2$ regularization on model parameters to prevent overfitting. The set $D=\{(u,i,j)|u\in \mathcal{U}, i\in \mathcal{I}_u^+,  j\in \mathcal{I} \setminus \mathcal{I}_u^+ \}$ denotes all pairwise training instances, where $\mathcal{U}$, $\mathcal{I}$, and $\mathcal{I}_u^+$ denote all users, items, and the interacted items of user $u$. To handle the sheer number of pairwise training instances, Rendle \etal~\cite{rendle2009bpr} advocate the use of stochastic gradient descent (SGD) for optimization, which is much less costly and converges faster than batch gradient descent. 

\subsubsection{Vulnerability of VBPR}\label{subsec:vul} Despite a sound solution for multimedia recommendation, we argue that VBPR is not robust in predicting user preference.
As demonstrated in Figure~\ref{fig:examples}, even small pixel-level perturbations on image candidates can yield large changes on the ranking of the candidates, which is out of expectation. Note that an image $i$ is converted to feature vector $\textbf{c}_i$ by DNN and the predictive model uses $\textbf{c}_i$ to predict user preference on the image (i.e., the $\textbf{h}_u^T (\textbf{E}\textbf{c}_i)$ term).
As such, it implies that two possibilities for the vulnerability of VBPR:  1) 
the small pixel-level changes result in large change on $\textbf{c}_i$, which subsequently leads to large change on the prediction value, and 2) the small pixel-level changes result in small changes on $\textbf{c}_i$, but even small fluctuations on $\textbf{c}_i$ can significantly change the prediction value. 

It is worth noting that both possibilities could be valid (e.g., exist for different instances) and can be supported by existing works. For example, Goodfellow \textit{et al.}~\cite{AML_2015} show that many DNN models are not robust to pixel-level perturbations (which provides evidence for the first possibility), and He \textit{et al.}~\cite{he2018adversarial} show that the MF model is not robust to purposeful perturbations on user and item embeddings (which provides evidence for the second possibility). Regardless of which exact reason, it points to the weak generalization ability of the overall multimedia recommender system --- if we imagine the prediction function as a curve in high-dimensional space, we can deduce that the curve is not smooth and has big fluctuations at many points. We believe that the vulnerability issue also exists for other deep feature-based multimedia recommendation methods, if no special action is taken to address the issue in the method. 
In this work, we address this universal issue in multimedia recommender systems by performing adversarial learning, which to our knowledge has not been explored before. 


%% file: 3_method.tex
\section{Adversarial Multimedia Recommendation} \label{sec:method}
This section elaborates our proposed method. We first present the predictive model, followed by the adversarial loss function, optimizing which can lead to a more robust recommender model. Lastly, we present the optimization algorithm. 

\subsection{Predictive Model}
\label{subsec:predM}
Note that the focus of this work is to train robust models for multimedia recommendation, rather than developing new predictive models. As such, we simply apply the model of VBPR and make slight adjustments on it:
\begin{equation}
\hat{y}_{ui} = \textbf{p}_u^T (\textbf{q}_i + \textbf{E}\cdot \textbf{c}_i),
\end{equation}
where $\textbf{p}_u\in \mathbb{R}^K$, $\textbf{q}_i\in \mathbb{R}^K$, $\textbf{E}\in \mathbb{R}^{K\times D}$ and $\textbf{c}_i\in\mathbb{R}^{D}$ have the same meaning as that in Equation (\ref{eq:vbpr}). The difference of this visually-aware recommender model with VBPR is that it associates each user with one embedding vector $\textbf{p}_u$ only, while in VBPR each user has two embedding vectors $\textbf{p}_u$
and $\textbf{h}_u$. This simplification is just to ensure a fair comparison with the conventional MF model when the embedding size $K$ is set as a same number (\textit{i.e.}, making the models have the same representation ability). Moreover, we have experimented with both ways of user embedding, and did not observe significant difference between them.



\subsection{Objective Function}
\label{subsec:objf}
Several recent efforts have shown that adversarial training can improve the robustness of machine learning models~\cite{AML_2015,miyato2017adversarial,he2018adversarial}. Inspired by their success, we develop adversarial training method to improve multimedia recommender model. The basic ideas are two-fold: 1) constructing an adversary that degrades model performance by adding perturbations on model inputs (and/or parameters), and meanwhile 2) training the model to perform well under the affect of adversary. In what follows, we describe the two ingredients of AMR's training objective function, namely, how to construct the adversary and how to learn model parameters. 
 \vspace{+5pt}

\begin{figure}
	\includegraphics[width=\linewidth]{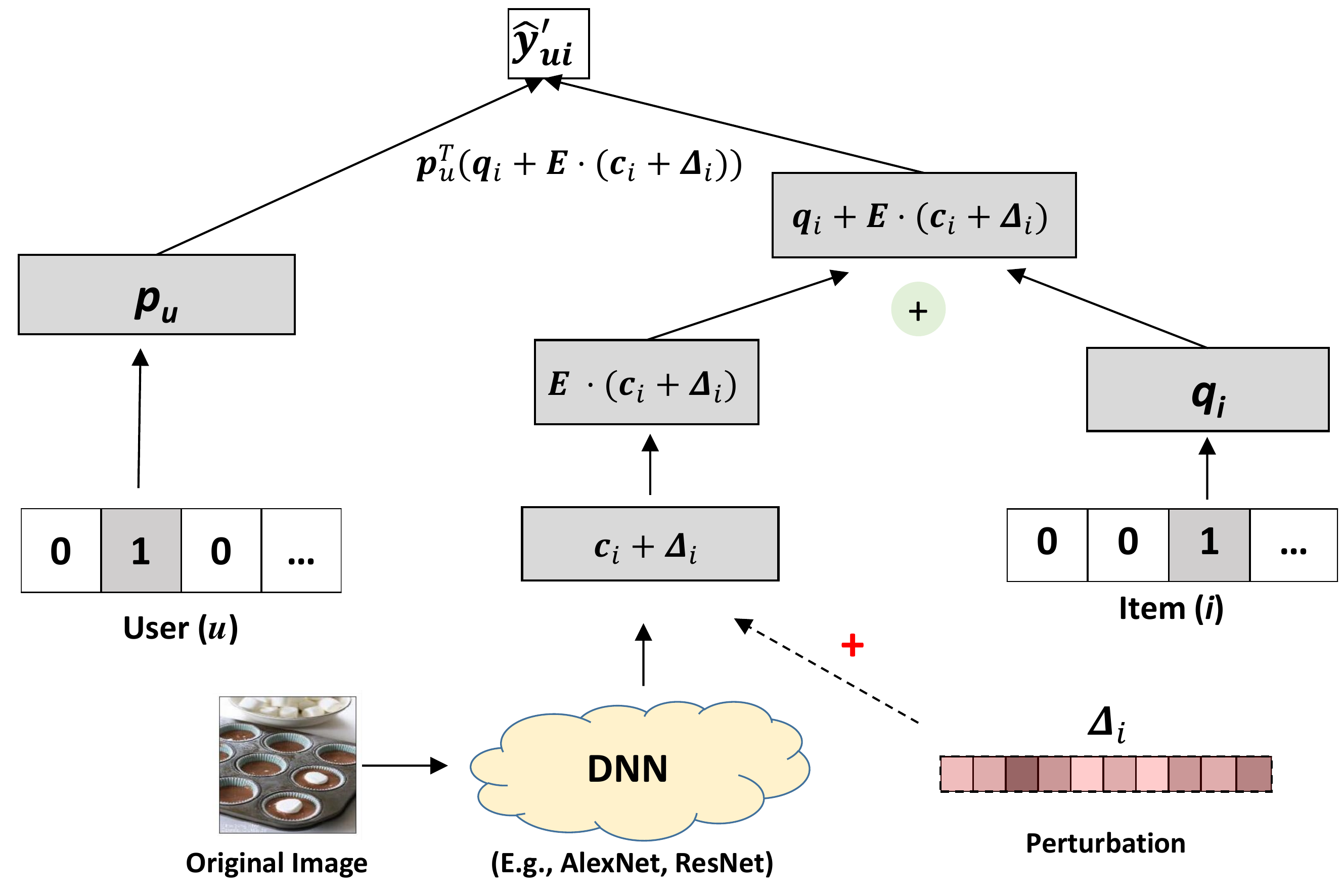}
	\caption{An illustration of the predictive model with perturbation $\Delta_i$, which is enforced on the image's feature vector extracted by DNN.}\label{fig:model}\vspace{-10pt}
\end{figure}

\noindent\textbf{1. Adversary Construction}. The goal of the constructed adversary is to decrease the model's performance as much as possible. Typically, additive perturbations are applied to either model inputs~\cite{AML_2015} or parameters~\cite{he2018adversarial}. 
To address the vulnerability issue illustrated in Figure~\ref{fig:examples}, an intuitive solution is to apply perturbations to model inputs, \textit{i.e.,} the raw pixels of the image, since the unexpected change on ranking result is caused by the perturbations on image pixels. Through this way, training the model to be robust to adversarial perturbations can increase the robustness of both the DNN (that extracts image deep features) and LFM (that predicts user preference). However, this solution is difficult to implement due to two practical reasons: 

First, it requires the whole system to be end-to-end trainable; in other words, the DNN for image feature extraction needs to be updated during the training of recommender model. Since user-item interaction data is sparse by nature and the DNN usually has many parameters, it may easily lead to overfitting issue if we train the DNN simutaneously. 

Second, it leads to a much higher learning complexity. Given a training instance $(u,i)$, the recommender model part only needs to update two embedding vectors ($\textbf{p}_u$ and $\textbf{q}_i$) and the feature transformation matrix $\textbf{E}$, whereas the DNN model needs to update the whole deep network, for which the parameters are several magnitudes larger. Moreover, to update the perturbations, we need to back-propagate the gradient through the DNN, which is also very time-consuming.

To avoid the difficulties in applying pixel-level perturbations, we instead propose to apply perturbations to the image's deep feature vector, \textit{i.e.,} $\textbf{c}_i$. To be specific, the perturbed model is formulated as:
\begin{equation}
\hat{y}_{ui}' = \textbf{p}_u^T (\textbf{q}_i + \textbf{E}\cdot (\textbf{c}_i + \Delta_i)),
\end{equation}
where $\Delta_i$ denotes the perturbations added on deep image feature vector by the adversary. Figure~\ref{fig:model} illustrates the perturbed model. This way of adding perturbations has two implications: 1) the DNN model can only serve as an image feature extractor, which is neither updated nor involved in the adversary construction process, making the learning algorithm more efficient, and 2) adversarial training can't improve the quality of deep image representation $\textbf{c}_i$, but it can improve the image's representation in MF's latent space (that is $\textbf{E}\textbf{c}_i$, since $\textbf{E}$ is updated by adversarial training towards the aim of being robust).  

We now consider how to find optimal perturbations that lead to the largest influence on the model, which are also known as the worst-case perturbations~\cite{AML_2015}. Since the model is trained to minimize the BPR loss (see Equation (\ref{eq:bpr})), a natural idea is to set an opposite goal for the perturbations --- maximizing the BPR loss. Let $\Delta = [\Delta_i] \in  \mathbb{R}^{|\mathcal{I}|\times D} $, which denotes the perturbations for all images and the $i$-th column is $\Delta_i$. We obtain optimal perturbations by maximizing the BPR loss on training data:
\begin{equation}
\begin{aligned}\label{eq:bpr_prime}
\Delta^* &= \arg\max_{\Delta} L'_{BPR} = \arg\max_{\Delta} \sum_{(u,i,j)\in \mathcal{D}} -\ln \sigma (\hat{y}_{ui}' - \hat{y}_{uj}'), \\ 
& \text{where}\quad ||\Delta_i|| \leq \epsilon, \quad \text{for}\quad  i = 1, ..., |\mathcal{I}|,
\end{aligned}
\end{equation}
where $||\cdot||$ denotes the $L_2$ norm, and $\epsilon$ is a hyper-parameter that controls the magnitude of perturbations. The constraint of $||\Delta_i|| \leq \epsilon$ is to avoid a trivial solution that increases the BPR loss by simply increasing the scale of $\Delta_i$.
Note that compared with the orignal BPR loss, we remove the $L_2$ regularizer on model parameters in this perturbed BPR loss, since the construction of $\Delta$ is based on the current values of model parameters, which are irrelevant to $\Delta$ and thus can be safely removed. \vspace{+5pt}

\noindent\textbf{2. Model Optimization}. To make the model less sensitive to the adversarial perturbations, in addition to minimize the original BPR loss, we also minimize the adversary's objective function. Let $\Theta$ be the model parameters, which includes $\textbf{p}_u$ for all users, $\textbf{q}_i$ for all items, and transformation matrix $\textbf{E}$. We define the optimization objective for the model as
\begin{equation}\label{eq:Theta}
\begin{aligned}
\Theta^* =& \arg\min_{\Theta} L_{BPR} + \lambda L_{BPR}', \\
=& \arg\min_{\Theta} \sum_{(u,i,j)\in \mathcal{D}} -\ln \sigma(\hat{y}_{ui} - \hat{y}_{uj}) - \lambda \ln \sigma (\hat{y}_{ui}' - \hat{y}_{uj}') \\
&+ \beta ||\Theta||^2  
\end{aligned}
\end{equation}
where $\lambda$ is a hyper-parameter to control the impact of the adversary on the model optimization. 
When $\lambda$ is set to 0, the adversary has no impact on training and the method degrades to VBPR. In this formulation, the adversary's loss $L_{BPR}'$ can be seen as regularizing~\cite{wang2016scalable} the model to make it be more robust, thus it is also called as \textit{adversarial regularizer} in literature~\cite{he2018adversarial}. \\

\noindent To unify the two processes, we formulate it as a minimax objective function. The optimization of model parameters $\Theta$ is the minimizing player, and the construction of perturbations $\Delta$ is the maximizing player:
\begin{equation}
\label{eq:ddd}
\begin{aligned}
\Theta^*, \Delta^* &= \arg \min_{\Theta} \max_{\Delta} L_{BPR} (\Theta) + \lambda L_{BPR}' (\Theta, \Delta), \\
& \text{where}\quad ||\Delta_i|| \leq \epsilon, \quad \text{for}\quad  i = 1, ..., |\mathcal{I}|.
\end{aligned}
\end{equation}
Compared to VBPR, our AMR has two more hyper-parameters to be specified ---  $\epsilon$ and $\lambda$. Both hyper-parameters are crucial to recommendation performance and need to be carefully tuned. Particularly, too large values will make the model robust to adversarial perturbations but at the risk of destroying the training process, while too small values will limit the impact of the adversary and make no improvements on the model robustness and generalization. 

Besides the minimax objective function, we can achieve similar effect of improving model robustness by employing random perturbations. That is, optimizing the model to make it perform well under stochastic noises on parameters. However, this way is less effective than our maximizing player, as we empirically show in Figure \ref{fig:damage}. Moreover, another option is to optimize only the adversarial regularizer $L'_{BPR}$. This manner may result in poor testing performance, as the model used for testing is the clean one without perturbations. Our minimax formulation can be understood as a way of data augmentation, which optimizes the model on both raw data and perturbed data simultaneously.
In the next subsection, we discuss how to optimize the minimax objective function. 

\begin{algorithm}[t]
	\caption{SGD learning algorithm for AMR.}
	\label{alg}
	\KwIn{Training data $\mathcal{D}$, adversarial noise level $\epsilon$, adversarial regularizer strength $\lambda$, $L_2$ regularizer strength $\beta$, and learning rate $\eta$;}
	\KwOut{Model parameters $\Theta$;}
	Initialize $\Theta$ from VBPR \;
	\While{not converge} {
		Randomly draw an example $(u,i,j)$ from $\mathcal{D}$ \;
		\tcp{Learning adversarial perturbations}
		$\Delta_i \leftarrow \epsilon \frac{\Gamma_i}{||\Gamma_i||}\quad \text{where} \quad \Gamma_i = \frac{\partial l_{uij}'}{\partial \Delta_i}$ \;
        $\Delta_j \leftarrow \epsilon \frac{\Gamma_j}{||\Gamma_j||}\quad \text{where} \quad \Gamma_j = \frac{\partial l_{uij}'}{\partial \Delta_j}$ \;
		\tcp{Learning model parameters}
		$\textbf{p}_u \leftarrow \textbf{p}_u - \eta \frac{\partial l_{uij}}{\partial \textbf{p}_u}$ \;
        $\textbf{q}_i \leftarrow \textbf{q}_i - \eta \frac{\partial l_{uij}}{\partial \textbf{q}_i}$ \;
        $\textbf{q}_j \leftarrow \textbf{q}_j - \eta \frac{\partial l_{uij}}{\partial \textbf{q}_j}$ \;
        $\textbf{E} \leftarrow \textbf{E} - \eta \frac{\partial l_{uij}}{\partial \textbf{E}}$ \;
	}
	\Return{$\Theta$}
\end{algorithm}

\subsection{Learning Algorithm}
Due to the large number of pairwise training instances in BPR loss, batch gradient descent could be very time consuming and slow to converge~\cite{rendle2009bpr}. As such, we prioritize the SGD learning algorithm. 
Algorithm~\ref{alg} illustrates our devised SGD learning algorithm for AMR. 
The subproblem to consider in SGD is that given a stochastic training instance $(u,i,j)$, how to optimize parameters related to this instance only (line 4-9): 

\noindent - For adversary construction (line 4-5), the objective function (maximized) regarding to this instance is:
\begin{equation}
l'_{uij} = -\ln \sigma(\hat{y}_{ui}' - \hat{y}_{uj}'),
\text{where}\ ||\Delta_i||\leq \epsilon,\  ||\Delta_j||\leq \epsilon.
\end{equation}

By \textbf{maximizing} the objective function, we obtain the worst-case perturbations $\Delta_i$ and $\Delta_j$, which can make the largest change on the BPR loss on the single instance $(u,i,j)$. 

\noindent - For model parameter learning (line 6-9), the objective function (to be minimized) regarding to this instance is:
\begin{equation}\label{eq:l_uai}
\begin{aligned}
l_{uij} =& - \ln \sigma(\hat{y}_{ui} - \hat{y}_{uj}) - \lambda \ln \sigma(\hat{y}_{ui}' - \hat{y}_{uj}') \\
&+ \beta (||\textbf{p}_u||^2 + ||\textbf{q}_i||^2  + ||\textbf{E}||^2).
\end{aligned}
\end{equation}

By \textbf{minimizing} the objective function, we obtain the model parameters $\textbf{p}_u, \textbf{q}_i, \textbf{E}$, which can the model resistant to the adversarial perturbations on the instance $(u,i,j)$. \vspace{+10pt}


\noindent In the next, we elaborate how to perform the two optimization procedures for a stochastic instance $(u,i,j)$.

\noindent \textbf{1. Learning Adversarial Perturbations}. This step obtains perturbed vectors that are relevant to model updates for the instance $(u,i,j)$, that is $\Delta_i$ and $\Delta_j$. 
Due to the non-linearity of the objective function $l_{uij}'$ and the $\epsilon$-constraint in optimization, it is difficult to get the exact solution.
As such, we borrow the idea from the fast gradient sign method~\cite{AML_2015}, approximating the objective function by linearizing it around $\Delta_i$ and $\Delta_j$; and then, we solve the constrained optimization problem on this approximated linear function.
According to Taylor series, the linear function is the first-order Taylor expansion, for which the line's slope is the first-order derivative of the objective function on the variables. 
It is clear that to maximize a linear function, the optimal solution is to move the variables towards the direction of their gradients. 
Taking the $\epsilon$-constraint into account, we can obtain the solution for adversarial perturbations as
\begin{equation}\label{eq:delta_i}
	\Delta_i = \epsilon \frac{\Gamma_i}{||\Gamma_i||}\quad \text{where} \quad \Gamma_i = \frac{\partial l_{uij}'}{\partial \Delta_i},
\end{equation}
\begin{equation}\label{eq:delta_j}
	\Delta_j = \epsilon \frac{\Gamma_j}{||\Gamma_j||}\quad \text{where} \quad \Gamma_j = \frac{\partial l_{uij}'}{\partial \Delta_j},
\end{equation}
Note that when a mini-batch of examples are sampled, $l_{uij}'$ should be defined as the sum of loss over the examples in the mini-batch, since the target item $i$ may also appear in other examples. Here we have omitted the details for the derivation, because modern machine learning toolkits like TensorFlow and PyTorch provide the auto-differential functionality. Moreover, we have also tried the fast gradient sign method as proposed in \cite{AML_2015}, which only keeps the sign of the derivation, \ie $\Delta_i = \epsilon \text{sign}(\Gamma_i)$. However, we find it is less effective than our solution on recommendation tasks. \vspace{+5pt}

\noindent\textbf{2. Learning Model Parameters}. This step updates model parameters by minimizing Equation (\ref{eq:l_uai}). Since the perturbations $\Delta$ are fixed in this step, it becomes a conventional minimization problem and can be approached with gradient descent. Specifically, we perform a gradient step for each involved parameter:
\begin{equation}\label{eq:sgd}
\theta = \theta - \eta \frac{\partial l_{uij}}{\partial \theta},
\end{equation}
where $\theta = \{\textbf{p}_u, \textbf{q}_i, \textbf{q}_j, \textbf{E}\}$. $\eta$ denotes the learning rate, which can be parameter-dependent if adaptive SGD methods are used, and we use the Adagrad~\cite{Adagrad} in our experiments. \vspace{+5pt}

\noindent For convergence, one can either check the decrease of $L_{BPR}$ after a training epoch (defined as iterating $|\mathcal{I}^+|$ number of examples where $|\mathcal{I}^+|$ denotes the number of observed interactions in the dataset), or monitor the recommendation performance based on a holdout validation set. 

Lastly, it is worth mentioning that the pre-training step (line 1 of Algorithm \ref{alg}) is critical and indispensable for AMR. This is because that only when the model has achieved reasonable performance, the model's generalization can be improved by enhancing its robustness with perturbations; otherwise, normal training process is sufficient to lead to better parameters and adversarial training will negatively slow down the convergence.

\subsection{Time Complexity Analysis}
We analyze the time complexity our AMR, with VBPR as a contrast. Since AMR and VBPR employ the same prediction model (the difference is in the training loss), they have the same time complexity in model prediction.

In order to better express the time complexity during training, let $O_f$ be the time complexity of forward propagation for $\hat{y}_{ui}$, $O_b$ be that of backward propagation, and $O_u$ be that of updating parameters. To compute the perturbations, there is an extra cost $O_{adv}$ to obtain the gradients on the content feature. According to the definitions, VBPR costs two $O_f$, two $O_b$ and one $O_u$ because of the pair-wised loss. That is totally $2\times O_f + 2\times O_b + O_u$. AMR does $O_f$ $O_f$ $O_{adv}$ $O_{adv}$ $O_f$ $O_f$ $O_b$ $O_b$ $O_b$ $O_b$ $O_u$ in sequence which is totally $4\times O_f + 4\times O_b + 2\times O_{adv} + O_u$. Usually, the four operations are linearly correlated. For example, their time complexities in VBPR and AMR are all $O(K+KD)$ (the divided $O(K)$ indicates the original part of MF). Thus the complexity of AMR is about two times of that of VBPR. Empirically, the time training AMR for one epoch is about three times of that training VBPR on both datasets. The redundant time cost may be led by some constant level operations. 

%% file: 5_experiment.tex
\section{Experiments} \label{sec:experiment}

In this section, we conduct experiments with the aim of answering the following questions:
\begin{description}
	\item{\textbf{RQ1}} Can our proposed AMR outperform the state-of-the-art multimedia recommendation methods?
	\item{\textbf{RQ2}} How is the effect of the adversarial training and can it improve the generalization and robustness of the model?
	\item{\textbf{RQ3}} How do the key hyper-parameters $\epsilon$ and $\lambda$ affect the performance?
\end{description}
We first describe the experimental settings, followed by results answering the above research questions. 

\subsection{Experimental Settings}

\subsubsection{Data Descriptions} We conduct experiments on two real-world datasets: Pinterest~\cite{ICCV15} and Amazon~\cite{VBPR}. 
On both datasets, 1) each item is associated with one image; and 2) the user-item interaction matrix is highly sparse. Table~\ref{tab:statistic} summarizes the statistics of the two datasets. 

\textbf{Pinterest}. The Pinterest data is used for evaluating the image recommendation task. Since the original data is extremely large (over one million users and ten million images), we sample a small subset of the data to verify our method. Specifically, we randomly select ten thousand users, and then discard users with less than two interactions and items without interactions. 

\textbf{Amazon}. The Amazon data is constructed by \cite{mcauley2015image} for visually-aware product recommendation. We use the \textit{women} category for evaluation. 
Similar to Pinterest, we first discard users with less than  five interactions. We then remove items that have no  interactions and correlated images.

\begin{table}
	\caption{Statistics of our experimented data.}
	\label{tab:statistic}
    \centering
	\vspace{-5pt}
	\begin{tabular}{|l|c|c|c|c|}
		\hline
		\textbf{Dataset} & User\# & Item\# & Interaction\# & Sparsity\\\hline
		Pinterest & 3,226 & 4,998 & 9,844 & 99.939\% \\\hline
		Amazon & 83,337 & 299,555 & 706,949 & 99.997\%\\\hline
	\end{tabular}
\end{table}

\subsubsection{Evaluation Protocol}
Following the prominent work in recommendation~\cite{rendle2009bpr,NCF}, we employ the standard \textit{leave-one-out} protocol. Specifically, for each user we randomly select one interaction for testing, and utilize the remaining data for training.  
After splitting, we find that about 52.6\%
and 45.9\% items in the testing set on Pinterest and Amazon respectively are cold-start (\ie out-of-sample) items. This poses challenges to traditional collaborative filtering methods and highlights the necessity of doing content-based filtering.
During training, these cold-start items are not involved (note that they can not be used as negative samples to avoid information leak); during testing, we initialize the ID embedding of cold-start items as a zero vector, using only their image features to get the item embedding.

Since it is time-consuming to rank all items for every user during evaluation, we follow the common approach \cite{elkahky2015multi,NCF} to sample 999 items that are not interacted with the user, and then rank the testing item among the 999 items.
To evaluate the performance of top-$N$ recommendation, we truncate the ranking list of the 1,000 items at position $N$, measuring its quality with the \textit{Hit Ratio}~(HR) and \textit{Normalized Discounted Cumulative Gain}~(NDCG). To be specific, HR@N measures whether the testing item occurs in the top-$N$ list --- 1 for yes and 0 for no; NDCG@N measures the position of the testing item in the top-$N$ list, the higher the better. 
The default setting of $N$ is 10 without special mention. 
We report the average scores of all users and perform one-sample paired t-test to judge the statistical significance when necessary.

\subsubsection{Baselines} We compare AMR with the following methods. 

\textbf{POP} is a non-personalized method that ranks items by their popularity, measured by the number of interactions in the training data. It benchmarks the performance of personalized recommendation.

\textbf{MF-eALS}~\cite{he2016fast} is a CF method that trains the MF model with a weighted regression loss, where different missing entries are assigned to different weights (i.e., confidence to be true negatives). It eschews negative sampling by assigning a uniform target of 0 on all missing entries, which allows fast optimization on MF. 

\textbf{MF-BPR}~\cite{rendle2009bpr} is a CF method that trains the MF model with BPR pairwise ranking loss. Since MF is learned solely based on user-item interactions, it serves as a benchmark for models with visual signals. 

\textbf{DUIF}~\cite{ICCV15} is a variant of LFM. It replaces the item embedding in MF by the projecting the deep image feature into the latent space. For a fair comparison with other methods, we also optimize DUIF with the BPR loss. We have tested DUIF by both training it from scratch and pre-training it with user embeddings of MF, and report the best results. 

\textbf{VBPR}~\cite{VBPR} is an extension of MF-BPR, which is tailored for visually-aware recommendation. The detailed description can be found in Section~\ref{subsec:vul}. For model initialization, we find that using the ID embeddings learned by MF leads to better performance, so we report this specific setting. 

Our \textbf{AMR} method is implemented using Tensorflow, which is available at: \url{https://github.com/duxy-me/AMR}. For visually-aware methods (DUIF, VBPR and AMR), we use the same ResNet-50~\cite{he2016deep} model\footnote{\url{https://github.com/KaimingHe/deep-residual-networks}} as the deep image feature extractor to make the comparison fair. Moreover, all models are optimized using mini-batch Adagrad with a mini-batch size of 512, and other hyper-parameters have been fairly tuned as follows. 

\subsubsection{Hyper-parameters Settings}
To explore the hyper-parameter space for all methods, we randomly holdout a training interaction for each user as the validation set. 
We fix the embedding size to 64 and tune other hyper-parameters as follows.
First, for baseline models MF-BPR, DUIF, and VBPR, we tune the learning rate in $[0.005, 0.01, 0.05, 0.1]$ and the $L_2$ regularizer in $[0, 10^{-6}, 10^{-4},$ $10^{-2}, 1]$.
After obtaining the optimal values of learning rate and $L_2$ regularizer for VBPR, we use them for our method and then tune the adversary-related hyper-parameters: $\epsilon$ and $\lambda$. Specifically, we first fix $\lambda=1$, tuning $\epsilon$ in $[10^{-2}, 10^{-1}, 1, 10]$. Then, with the best $\epsilon$, we tune $\lambda$ in $[10^{-2}, 10^{-1}, 1, 10]$. Note that if the optimal value was found in the boundary, we further extend the boundary to explore the optimal setting. We report the best results for all methods. 

\subsection{Performance Comparison (RQ1)}
\begin{table*}[t]
	\centering
	\caption{Top-$N$ recommendation performance where $N\in\{5,10,20\}$. RI is the relative improvement of AMR over baselines on average. $^*$ indicates that the improvements over baselines are statistically significant for $p<0.05$.} \vspace{-5pt}
	\label{tab:performance}
	\resizebox{\linewidth}{!}{
	\begin{tabular}{|l|c|c|c|c|c|c|c||c|c|c|c|c|c|c|}
		\hline
		&  \multicolumn{7}{c||}{\textbf{Pinterest}} & \multicolumn{7}{c|}{\textbf{Amazon}} \\\hline
		& \multicolumn{3}{c|}{\textbf{HR@$N$}} & \multicolumn{3}{c|}{\textbf{NDCG@$N$}} & RI & \multicolumn{3}{c|}{\textbf{HR@$N$}} & \multicolumn{3}{c|}{\textbf{NDCG@$N$}} & RI\\\hline
               & $N=5$  & $N=10$ & $N=20$ & $N=5$  & $N=10$ & $N=20$ &        & $N=5$  & $N=10$ & $N=20$ & $N=5$  & $N=10$ & $N=20$ & \\\hline\hline
	Pop		&	0.0353	&	0.0604	&	0.0927	&	0.0213	&	0.0296	&	0.0376	&	281.73\%
			&	0.1003	&	0.1460	&	0.2040	&	0.0685	&	0.0832	&	0.0978	&	34.27\%\\\hline
	
	MF-eALS    &   0.1262 & 0.1584 & 0.1953 & 0.0922 & 0.1025 & 0.1118 & 22.18\%
	        & 0.1322 & 0.1660 & 0.2026 & 0.1009 & 0.1118 & 0.1211 & 7.98\%\\\hline
	
	MF-BPR	&	0.1228	&	0.1534	&	0.1891	&	0.0949	&	0.1048	&	0.1138	&	22.82\%
			&	0.1306	&	0.1720	&	0.2183	&	0.0950	&	0.1084	&	0.1201	&	7.91\%\\\hline
	DUIF	&	0.1116	&	0.1600	&	0.2179	&	0.0806	&	0.0962	&	0.1108	&	26.17\%
			&	0.0865	&	0.1317	&	0.1964	&	0.0568	&	0.0714	&	0.0876	&	52.61\%\\\hline
    VBPR	&	0.1352	&	0.1829	&	0.2347	&	0.1005	&	0.1157	&	0.1287	&	7.70\%
			&	0.1333	&	0.1747	&	0.2249	&	0.0980	&	0.1113	&	0.1240	&	5.14\%\\\hline
	AMR		&	\textbf{0.1392}$^*$	&	\textbf{0.2033}$^*$	&	\textbf{0.2697}$^*$	&	\textbf{0.1026}$^*$	&	\textbf{0.1230}$^*$	&	\textbf{0.1398}$^*$	&	-	
			&	\textbf{0.1402}	&	\textbf{0.1864}$^*$	&	\textbf{0.2360}$^*$	&	\textbf{0.1022}	&	\textbf{0.1171}$^*$	&	\textbf{0.1296}$^*$	&	-\\\hline
		
	\end{tabular}%
	}
\end{table*}
Here we compare the performance of our AMR with baselines. We explore the top-$N$ recommendation where $N \in \{5,10,20\}$. The results are listed in Table~\ref{tab:performance}. Inspecting the results from top to bottom, we have the following observations.


First, on both datasets, personalized models (\ie MF-eALS, MF-BPR, VBPR and AMR) largely outperform the non-personalized method POP. Particularly, the largest improvements can achieve 280\% on Pinterest as indicated by the RI column. This demonstrates the positive effect of doing personalization. 

Second, among the personalized methods, VBPR outperforms MF-eALS, MF-BPR and DUIF in most cases. 
The improvements of VBPR over MF-BPR confirm that traditional CF models can be significantly enhanced by adding rich multimedia features. Meanwhile, we notice that DUIF shows much worse results than MF-BPR and MF-eALS even it has used the same visual features as VBPR. 
Considering the fact that DUIF leverages the multimedia features only to represent an item, we speculate that CF features (\ie ID embeddings) are more important than pure multimedia features in personalized recommendation.

Third, AMR consistently outperforms all baselines in terms of all metrics on both datasets. One advantage is that AMR is built on VBPR which performs better than BPR in general. More importantly, by introducing the adversarial examples in the training phase, AMR can learn better model parameters than the non-adversarial VBPR. Moreover, we test the performances on cold items, i.e., testing items that are not occurred in the training set. In this case, only image features are used to obtain the item embedding vector. 
In Pinterest, the HR@10/NDCG@10 improvement over VBPR is 74\%/101\%; in contrast, on non-cold items, the HR@10/NDCG@10 improvement is 6.2\%/ 2.2\%. This justifies the positive effect of our AMR on learning better image representations, thus leads to better multimedia recommendation performance, particularly for cold items are heavily relied on the image representations. 

Finally, focusing on Amazon, we find that the improvements of MF-BPR over POP, VBPR over MF-BPR, and AMR over VBPR, are smaller than that on the Pinterest data. The reasons lie in several aspects. First, the relatively strong performance of POP indicates that popular products on Amazon are more likely to be purchased by users; by contrast, the click behaviors on Pinterest images do not exhibit such pattern. 
Second, the small improvements of VBPR over MF-BPR reveal that adding multimedia content features have only minor benefits when the CF effect is strong (evidenced by richer user-item interactions, see Table \ref{tab:statistic} for more details). That may explain why multimedia information is typically regarded as an auxiliary but not dominant feature in recommender system domain. 
Therefore, the result that AMR has smaller improvements over VBPR is acceptable, since on the Amazon data, the recommendation quality is not dominant by visual features and thus modeling them only have minor effects. 
Despite this, by using adversarial training, our AMR can still improve over VBPR significantly, as evidenced by the t-test. This demonstrates the usefulness of adversarial training in improving the overall generalization of the model.

\subsection{Effect of Adversarial Training (RQ2)}
In this subsection, we analyze the effect of adversarial training from two aspects: generalization and robustness. 

\subsubsection{Generalization} 
\begin{figure}
	\includegraphics[width=\linewidth]{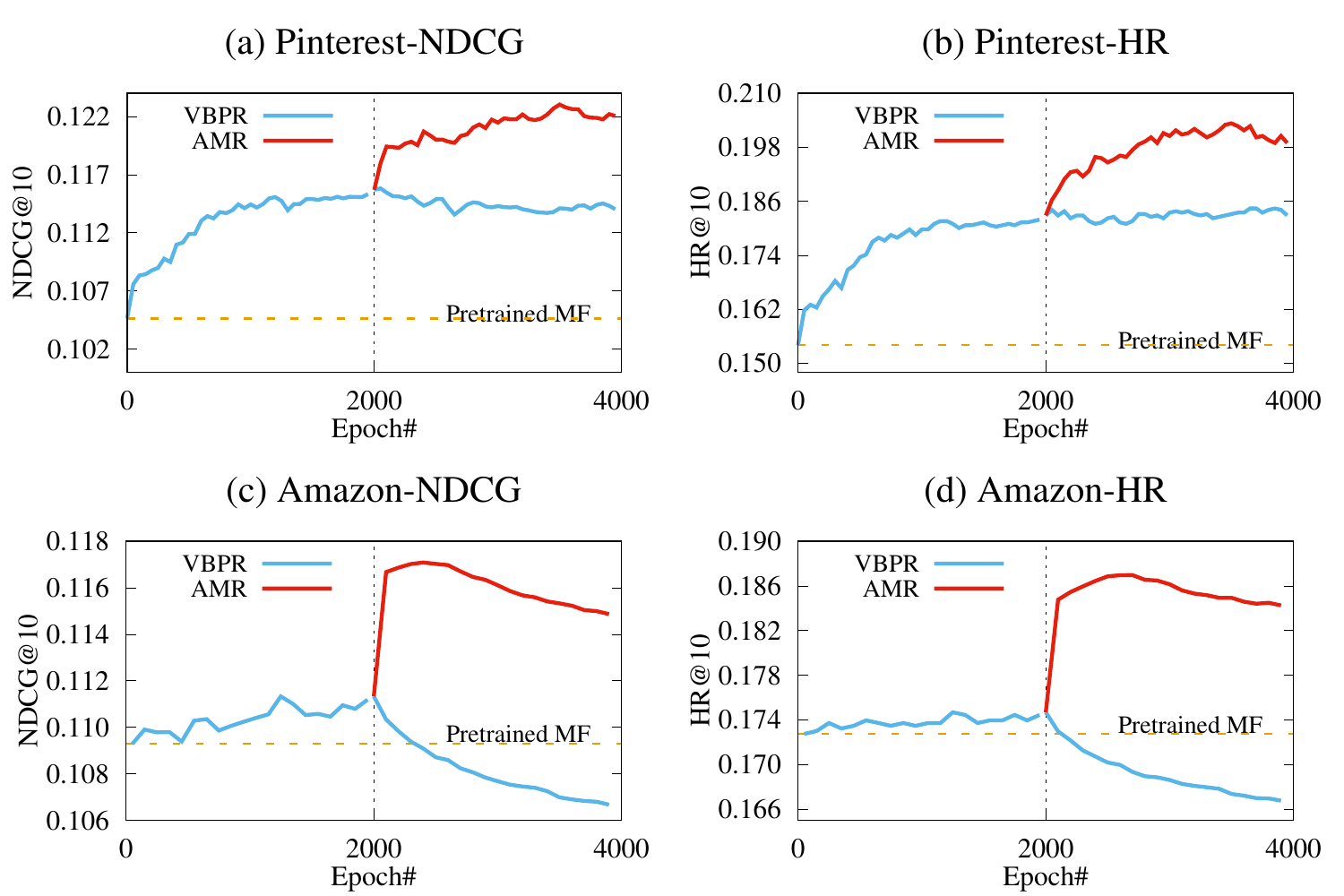}
	\caption{Testing performance of VBPR and AMR evaluated per 50 epochs. We first train VBPR for 2000 epochs (which is initialized from MF parameters for better performance). We then continue training AMR for another 2000 epochs (with continue training VBPR as a comparison).}
  \label{fig:train}
\end{figure}

We show the training process of VBPR and AMR in Figure~\ref{fig:train}, where the $y$-axis denotes the testing performance evaluated per 50 epochs. We also show the performance of pretrained MF as a benchmark, since VBPR and AMR are initialized from MF parameters.
Specifically, we first train VBPR until convergence (about 2000 epochs). Then we proceed to train AMR by initializing its parameters with the parameters learned by VBPR. As a comparison, we use the same parameters to initialize a new VBPR model and continue training it. As can be seen, by performing adversarial training based on VBPR parameters, we can gradually improve the performance to a large extent. By contrast, when performing normal training on VBPR, the performance is not improved, or even decreased due to overfitting (see results on Amazon).
To be specific, on the Pinterest dataset, the best NDCG and HR of VBPR are 0.116 and 0.183 respectively, which are further improved to 0.123 and 0.203 by training with AMR. 
These results verify the highly positive effect of adversarial learning in AMR, which leads to better parameters that can improve model generalization.

\subsubsection{Robustness}

We now recap the motivating example about model robustness in Figure~\ref{fig:examples}. To have a quantitative sense on the model robustness, we add adversarial perturbations to the original image and measure performance drop; smaller drop ratio means stronger robustness.

\begin{figure}
	\includegraphics[width=\linewidth]{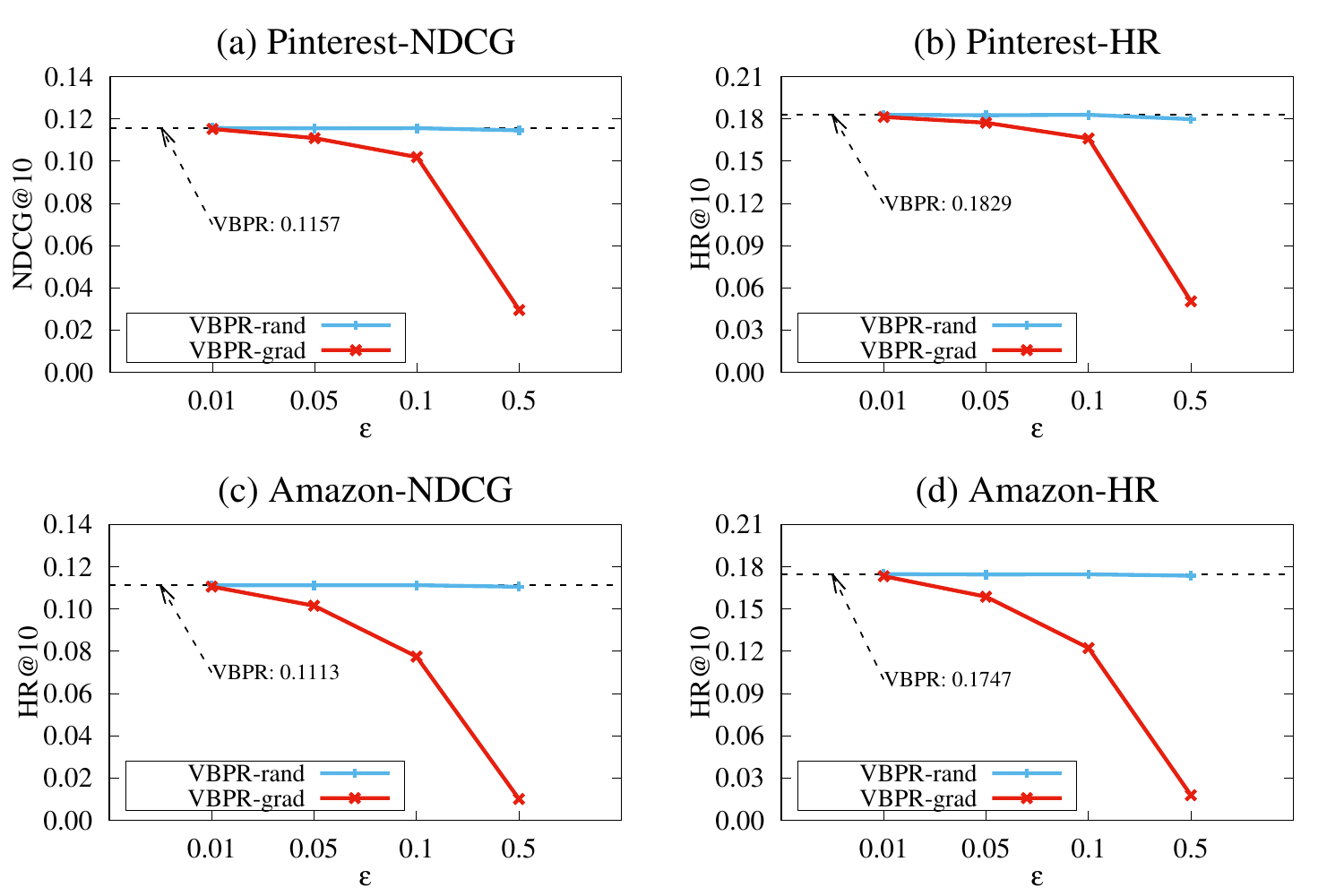}
	\caption{Impact of applying random (VBPR-rand) and adversarial (VBPR-grad) perturbations to image features on VBPR. The key observation is that adversarial perturbations have a large impact on BPR.}
    \label{fig:damage}
\end{figure}
We first demonstrate the impact of perturbations on VBPR. Figure~\ref{fig:damage} exhibits the performances. The horizontal dashed line indicates the performance of unperturbed VBPR. The random perturbation~(VBPR-rand) would decrease the performance by a small ratio. In contrast, our proposed adversarial perturbation~(VBPR-grad) introduced in Equation~\ref{eq:ddd} leads to a terrible impact on VBPR. Thus AMR addresses adversarial perturbation for the robust model. Another interesting observation is that the drop caused by the perturbation are exponential-like increased. The perturbation with $\epsilon = 0.01$ is inconspicuous, while the perturbation with $\epsilon = 0.5$ causes a fatal drop. That is a reference for setting $\epsilon$ in AMR.

\begin{table}
	\centering
	\caption{Performance drop (relatively decreasing ratio in NDCG@10) of VBPR and AMR in the presence of adversarial perturbations during the testing phase.}\vspace{-5pt}
	\label{tab:robust}
	\resizebox{\linewidth}{!}{
		\begin{tabular}{|l|c|c|c|c|c|c|}
			\hline
			&  \multicolumn{2}{c|}{$\epsilon=0.05$} &  \multicolumn{2}{c|}{$\epsilon=0.1$} &  \multicolumn{2}{c|}{$\epsilon=0.2$}\\\hline
			\textbf{Dataset} & \textbf{VBPR} & \textbf{AMR} & \textbf{VBPR} & \textbf{AMR} & \textbf{VBPR} & \textbf{AMR}\\\hline
			Pinterest & -4.2\% & -2.6\% & -11.9\% & -6.2\% & -31.8\% & -18.4\%\\\hline
			Amazon    &  -8.7\% &  -1.4\% & -30.4\% &  -5.3\% & -67.7\% & -20.2\%\\\hline
		\end{tabular}%
	}
\end{table}

Table~\ref{tab:robust} shows the relative performance drop of VBPR and AMR with different settings of $\epsilon$ (which controls the perturbation scale). We can see that across settings AMR has a much smaller performance drop than VBPR. For example, on Amazon, when $\epsilon$ sets to 0.05, VBPR decreases for $8.7\%$ whereas AMR decreases for $1.4\%$, which is about 6 times smaller. These results provide important empirical evidence on the robustness of AMR, which is less vulnerable to adversarial examples than VBPR. Moreover, larger perturbations, denoted by the increasing $\epsilon$, impact both models more severely. The perturbation with $\epsilon=0.2$ almost damage VBPR model since the performance drop of VBPR on Amazon is more than $67.7\%$. The drop of AMR is about $1/3$ of that of VBPR. These comparisons reveal the robustness of our proposed model AMR.

\begin{figure*}[t!]
    \centering
    \begin{subfigure}[t]{0.5\linewidth}
        \caption{Impact on Pinterest.}
        \centering
        \includegraphics[width=\linewidth]{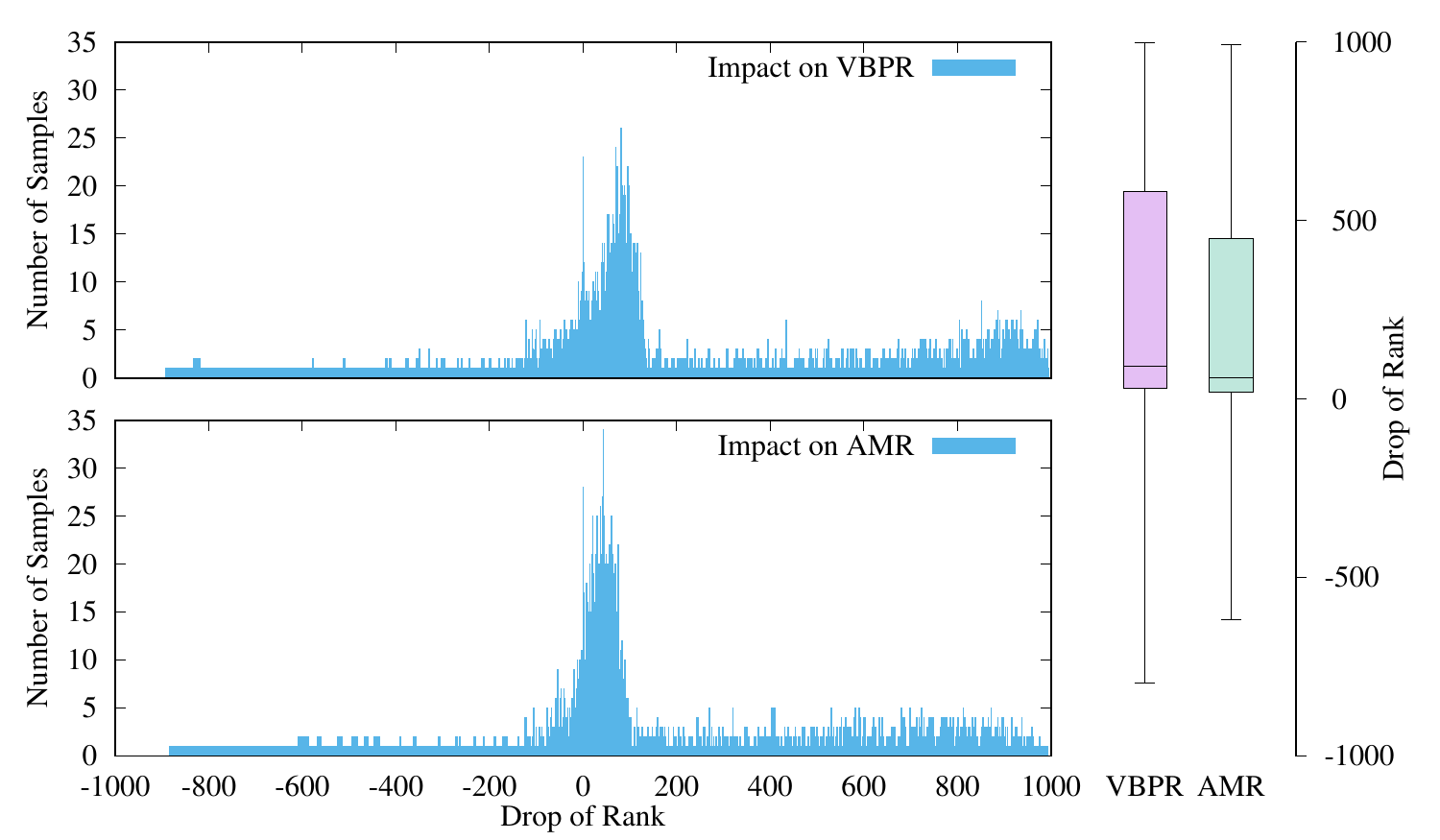}
    \end{subfigure}%
    ~ 
    \begin{subfigure}[t]{0.5\linewidth}
        \caption{Impact on Amazon.}
        \centering
        \includegraphics[width=\linewidth]{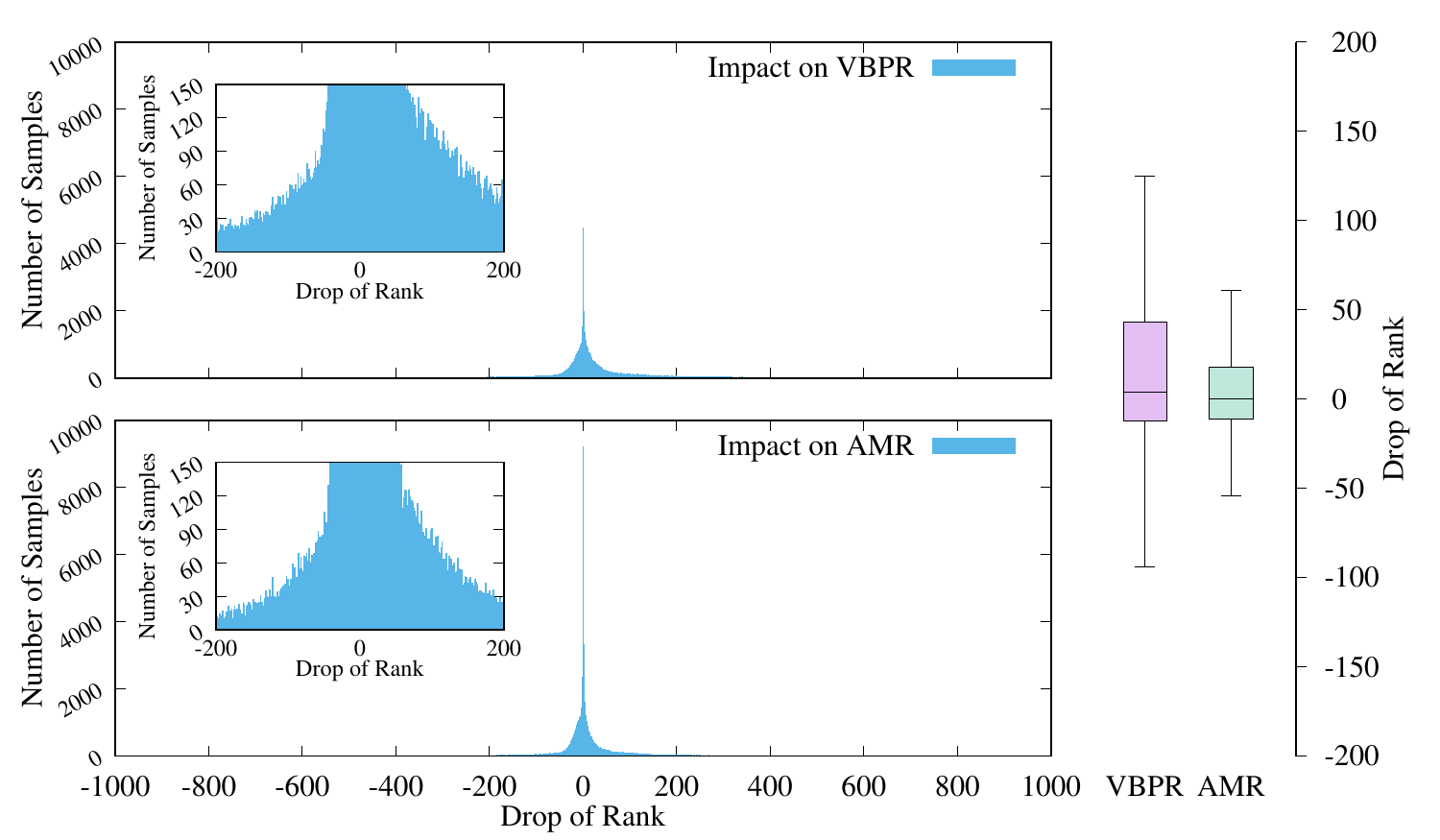}
    \end{subfigure}
    \caption{The impacts with perturbations under $\epsilon=0.1$. The boxplots represent the statistical distributions at the left side of them. The results, that the rank drop of AMR are closer to the zero point than that of VBPR, reveal that AMR gives more robust predictions than VBPR when facing perturbations.}
    \label{fig:dist}
\end{figure*}

We further explore the changes when applying the perturbations with $\epsilon = 0.1$. We record the subtraction of the sample ranks without and with perturbations. The positive values indicate the bad impact. $0$ means there are no changes. Figure~\ref{fig:dist} records the distribution of the subtractions. There are three major observations: 
\begin{enumerate}
    \item Most of the perturbations lead to worse performance, while a few lead to better performance. This is caused by the disorder of the predictions.
    \item On both Pinterest and Amazon, the large impact of AMR is fewer than that of VBPR, and the mean and the variance of the drops of AMR are less than those of VBPR. Specifically, on Amazon, most of the samples do not have any changes so that AMR would give a stable recommending results. These situations verify the robustness of AMR.
    \item There are a large ratio of changes larger than 500 in Pinterest while the ratios is much smaller in Amazon. That demonstrates that larger dataset may be relatively stable facing the perturbations. 
\end{enumerate}



\subsection{Hyper-parameter Exploration (RQ3)}
In this final subsection, we examine the impact of hyper-parameters of adversarial learning, \ie $\epsilon$ and $\lambda$, which control the scale of perturbation and the weight of adversary, respectively. In exploring the change of one hyper-parameter, all other hyper-parameters are fixed to the same (roughly optimal) value. 

\begin{figure}
	\includegraphics[width=\linewidth]{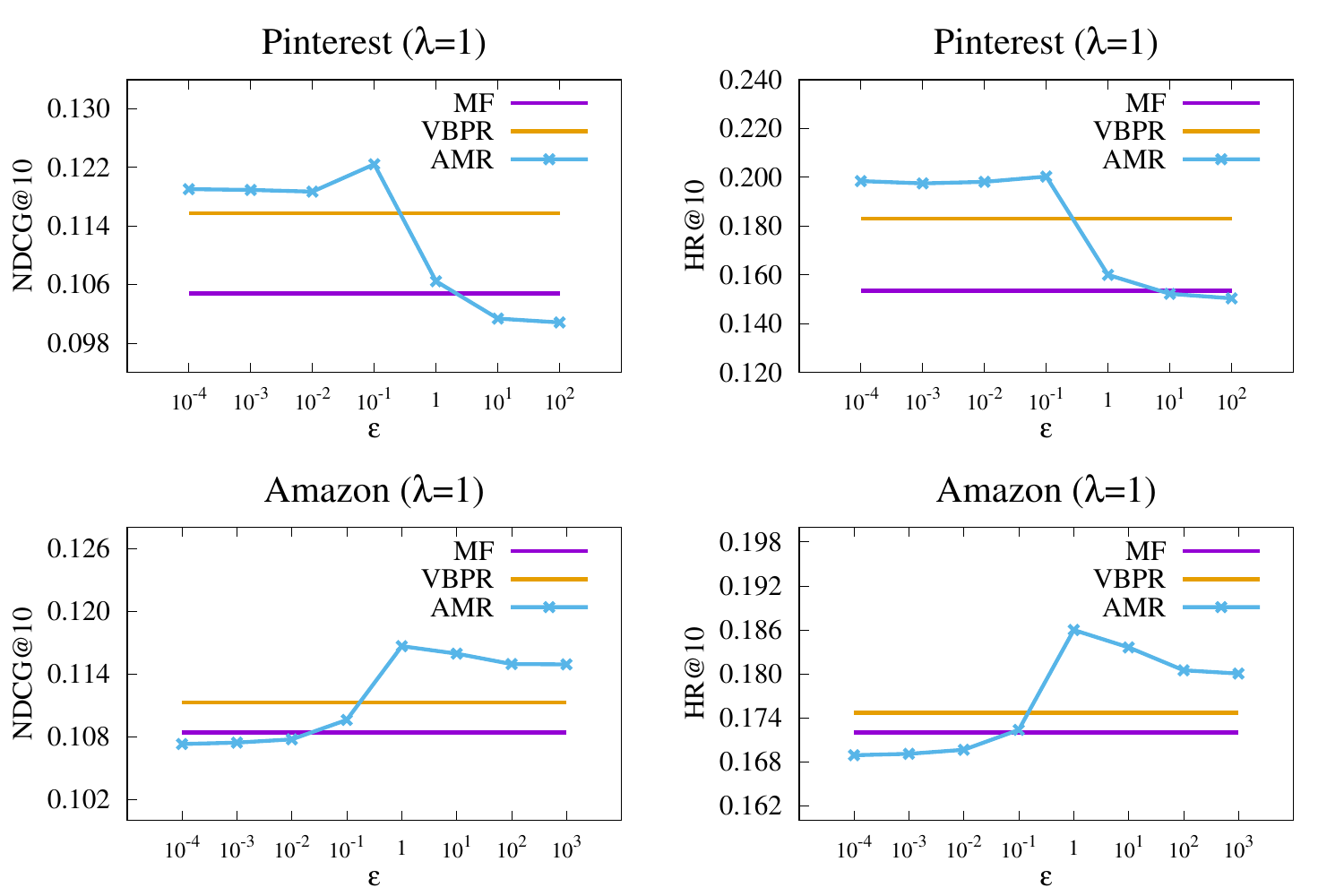}
	\caption{Performance of AMR \wrt different values of $\epsilon$. AMR obtains the best performance when $\epsilon=0.1$ and $\epsilon=1$ on Pinterest and Amazon, respectively.}
	\label{fig:eps}
\end{figure}

\begin{figure}
	\includegraphics[width=\linewidth]{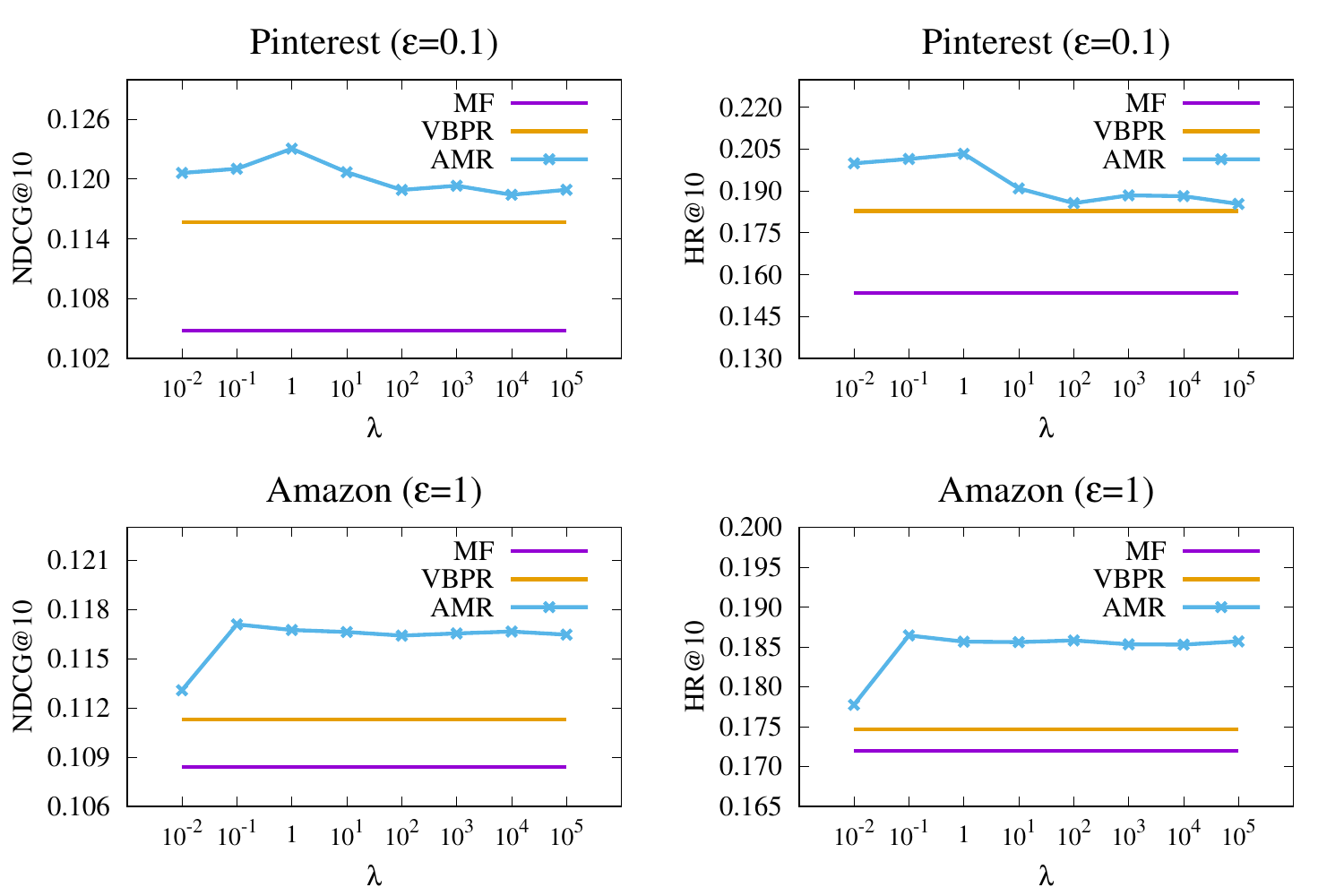}
	\caption{Performance of AMR \wrt different values of $\lambda$. AMR obtains the best performance with $\lambda=1$ and $\lambda=0.1$ on Pinterest and Amazon, respectively.}
	\label{fig:lmd}
\end{figure}

Figure~\ref{fig:eps} illustrates the performance change with respect to $\epsilon$. We can see that the optimal results are obtained when $\epsilon=0.1$ and $\epsilon=1$ on Pinterest and Amazon, respectively. When $\epsilon$ is smaller than 1, increasing it leads to gradual improvements. This implies the utility of adversarial training when the perturbations are within a controllable scale. However, when $\epsilon$ is larger than the optimal point, the performance of AMR drops rapidly, which reveals that too large perturbations will destroy the training process. 
Figure~\ref{fig:lmd} shows the results of varying $\lambda$. 
We can see that similar trends can be observed --- when $\lambda$ is smaller than a threshold, increasing it will improve the performance, and further increasing it beyond the threshold will decrease the performance significantly. Moreover, the threshold (\ie optimal $\lambda$) is different for the two datasets --- 1 for Pinterest and 0.1 for Amazon, which indicates that the optimal setting of $\lambda$ is data-independent and should be separately tuned for a dataset.



%% file: 4_related.tex
\section{Related Work} \label{sec:related}
In this section, we briefly review related work on  multimedia recommendation and adversarial learning. 

\subsection{Multimedia Recommendation}
In recommender system research, two lines of contributions are most significant to date: 1) pure Collaborative Filtering~(CF) techniques such as matrix factorization~\cite{fastMF} and its variants~\cite{NCF}, and 2) content- or context- aware methods that rely on more complex models such as feature-based embeddings~\cite{yuan2016lambdafm} and deep learning~\cite{NFM,AFM}. While multimedia recommendation falls into the second category of content-based recommendations, it is more challenging yet popular, due to massive and abundant multimedia (\eg visual, acoustic and semantic) features in real-world information systems~\cite{Oord:2013:DCM,kang2017visually,wang2018first}.

To effectively leverage rich multimedia features, a variety of multimedia recommendation techniques have been proposed. For example, it is intuitive to integrate high-level visual features that are extracted from DNNs into traditional CF models. A typical method is VBPR~\cite{VBPR} that extends the dot product-based embedding function in BPR~\cite{rendle2009bpr} into visual feature-based predictors. While simple, VBPR demonstrates considerable improvements in recommendation quality due to the proper use of multimedia features. Similarly, DUIF~\cite{ICCV15} builds item embedding by converting from the CNN feature of the image. Following the two works, Liu~\etal~\cite{liu2017deepstyle} takes the categories and styles annotated by CNNs as item features. Moreover,  Lei~\etal~\cite{lei2016comparative} and Kang~\etal~\cite{kang2017visually} do not directly use the features extracted in advance, but instead construct an end-to-end model by CNNs. At a finer granularity, Chen~\etal~\cite{chen2018visually} and ACF~\cite{ACF} crop images into several parts, and then integrate the features from each part with an attention mechanism, which has been an import technique in recommendation~\cite{TEM,cheng20183ncf}. Several other features have been exploited, such as acoustic ~\cite{deldjoo2018using,cheng2017exploiting}, aesthetic~\cite{yu2018aesthetic}, relation-based~\cite{xun2019trans} and location-aware features~\cite{zhao2017photo2trip,chen2018venue}.

The key idea of AMR is to increase the model robustness by making it less vulnerable to worst-case perturbations in input features. While the idea is originated from the two ICLR papers~\cite{AML_2014,AML_2015}, we are the first to implement the idea on multimedia recommender models and verify its effectiveness. Specifically, the two original ICLR papers worked on the classification task (which optimized the point-wise cross-entropy loss), whereas our work AMR addresses the ranking task (which optimized the pair-wise loss). Thus, for multimedia recommendation methods that optimize pair-wise ranking loss, such as the ACF~\cite{ACF}, our method can be directly applied; for methods that optimize pointwise loss, such as the Personalized Key-frame Recommendation~\cite{ACF} and Deep Content-based Music Recommendation~\cite{Oord:2013:DCM}, we just need to adapt the loss in model optimization (Equation~\ref{eq:Theta}), whereas the learning procedure remains unchanged.

\subsection{Adversarial Learning}
Another relevant line of research is adversarial learning~\cite{Lowd:2005}, which aims to find malicious examples to attack a machine learning method and then addresses the vulnerabilities of the method. Recent efforts have been intensively focused on DNNs owing to their extraordinary abilities in learning complex predictive functions. For example, Szegedy~\etal~\cite{AML_2014} finds that several state-of-the-art DNNs consistently mis-classify adversarial images, which are formed by adding small perturbations that maximize the model's prediction error. While the authors speculated that the reason is caused by the extreme nonlinearity of DNNs, later findings by Goodfellow~\etal~\cite{AML_2015} showed that the reason is opposite --- the vulnerability stems from the linearity of DNNs. They then proposed the fast gradient sign method that can efficiently generate adversarial examples with the linear assumption. Later on, the idea has been extended to several NLP tasks such as text classification~\cite{miyato2017adversarial}. Besides adding perturbations to input, other attempts have been made on the embedding layer~\cite{miyato2017adversarial} and dropout~\cite{AdversarialDropout}. 

In the domain of recommendation, there are very few efforts exploring the vulnerability of recommender models. Some previous work~\cite{book:robustCF} enhance the robustness of a recommender system by making it resistant to profile injection attacks, which try to insert fake user profiles to change the behavior of collaborative filtering algorithms. This line of research is orthogonal to this work, since we consider improving the robustness of recommender system from a different perspective of multimedia content.
The work that is most relevant with ours is~\cite{he2018adversarial}, which proposes a general adversarial learning framework for personalized ranking (aka., adversarial personalized ranking, short for APR). The key differences of AMR with APR are 1) APR is a general recommender framework focusing on the fundamental CF structure while AMR is a model focusing on multimedia recommendation with rich visual features, and 2) APR applies the perturbations on embeddings to increase the robustness of latent representations while AMR applies the perturbations on image features to increase the model tolerance for noisy inputs.
To the best of our knowledge, this is the first work that explores adversarial learning in multimedia recommendation, opening a new door of improving the robustness and generalization of multimedia recommender systems. 

%% file: 6_conclusion.tex
\section{Conclusion} \label{sec:conclusion}
In this work, we first showed that VBPR, a state-of-the-art image-aware recommendation method, is vulnerable to adversarial perturbations on images. The evidence is that by changing the images with very small perturbations that are imperceptible by human, we observed significant drop in recommendation accuracy. To address the lack of robustness issue of DNN-based multimedia recommender systems, we presented a new recommendation solution named AMR. By simultaneously training the model and the adversary that attacks the model with purposeful perturbations, AMR obtains better parameters, which not only make the model more robust but also more effective. Extensive results on two real-world datasets demonstrate the utility of adversarial learning and the strength of our method. 

In essence, AMR is a generic solution not limited to the model explored in this paper, but can serve as a general blueprint for improving any content-based recommender models.  In future, we plan to extend the AMR methodology to more models, such as the attention-based neural recommender models~\cite{ACF} which might be more effective than LFM. Moreover, we will incorporate more contexts for multimedia recommendation, such as time, location, and user personality. Lastly, we are interested in building interactive recommender systems by unifying the recent advances in dialog agents with recommendation technologies. 

\section*{Acknowledgement}
The authors thank the anonymous reviewers for their reviewing efforts. 
This work was partially supported by the National Key Research and Development Program of China under Grant 2016YFB1001001, the National Natural Science Foundation of China (Grant No. 61772275, 61732007, 61720106004 and U1611461), Linksure Network Holding Pte Ltd and the Asia Big Data Association (Award No.: AISG-100E-2018-002), and the Natural Science Foundation of Jiangsu Province (BK20170033). This research is also part of NExT++, supported by the National Research Foundation, Prime Ministers Office, Singapore under its IRC@Singapore Funding Initiative.